\documentclass{ws-ijmpd}

\begin{document}

\markboth{U. Mukhopadhyay, Saibal Ray and S. B. Dutta Choudhury}
{$\Lambda$-CDM Universe: A Phenomenological Approach With Many
Possibilities}

%
\catchline{}{}{}{}{}
%

\title{$\Lambda$-CDM Universe: A Phenomenological Approach With Many
Possibilities}

\author{UTPAL MUKHOPADHYAY}

\address{Satyabharati Vidyapith, North 24
Parganas, Kolkata 700 126, West Bengal, India.}

\author{SAIBAL RAY}

\address{Department of Physics, Barasat Government College, North 24 Parganas, Kolkata 700
124, West Bengal, India\footnote{Corresponding address}\\
saibal@iucaa.ernet.in}

\author{S. B. DUTTA CHOUDHURY}
  \address{Department of Physics, Jadavpur University,
  Kolkata 700 032, West Bengal, India}

\maketitle

\begin{history}
\received{Day Month Year} \revised{Day Month Year}
\end{history}

\begin{abstract}
A time-dependent phenomenological model of $\Lambda$, viz., $\dot
\Lambda\sim H^3$ is selected to investigate the $\Lambda$-CDM
cosmology. Time-dependent form of the equation of state parameter
$\omega$ is derived and it has been possible to obtain the sought
for flip of sign of the deceleration parameter $q$. Present age of
the Universe, calculated for some specific values of the
parameters agrees very well with the observational data.
\end{abstract}


\section{Introduction}
Modern cosmological research rests heavily on observational data.
Any theoretical model should be corroborated with observation for
understanding the viability of that model. Present cosmological
picture, emerging out of this theory-observation combination,
reveals that, the total energy-density of the Universe is
dominated by two dark components, viz., dark matter and dark
energy. Observational evidence from various independent sources
including SN
Ia\cite{Riess1998,Perlmutter1999,Knop2003,Spergel2003,Riess2004,Tegmark2004a,Astier2005,Spergel2007}
suggest that the cosmic expansion is speeding up, i.e.
accelerating. This acceleration is supposed to be caused by a yet
unknown exotic energy, termed as dark energy. A special feature of
dark energy is that it exerts negative pressure which acts as a
repulsive force initiating the observed acceleration. Not only
that, it is now well-accepted that about two-third of the total
energy-density comes in terms of dark energy while the remaining
one-third is contributed by matter, both visible and
dark\cite{Sahni2004}.

Now, dark matter played a significant role in the early Universe
during structure formation because it clumps in sub-megaparsec
scales. But, the exact composition of dark matter is still
unknown. Since small density perturbation ($\delta\rho/\rho\sim
10^{-5}$ at $z\simeq 1100$) measured by COBE and CMB
experiments\cite{Sahni2004} rule out baryonic dark matter and hot
dark matter like light neutrinos do not support hierarchical
structure formation
\cite{Elgaroy2002,Minakata2003,Spergel2003,Ellis2003}, so most of
the dark matter must be cold and non-baryonic. On the other hand,
clustering of cold dark matter on small scale\cite{Sahni2004}
supports the hierarchical structure formation. Moreover, after
introduction of the idea of accelerating Universe, the previous
Standard Cold Dark Matter (SCDM) models have fallen out of
grace\cite{Efstathiou1990,Pope2004} and is replaced by
$\Lambda$-CDM or LCDM model for including dark energy as a part of
the total energy density of the Universe. $\Lambda$-CDM model is
found to be in nice agreement with various sets of
observations\cite{Tegmark2004b}. An advantage of $\Lambda$-CDM
model is that it assumes a nearly scale-invariant primordial
perturbations and a Universe with no spatial curvature. These were
predicted by inflationary
scenario\cite{Mukhanov1981,Guth1982,Hawking1982,Starobinsky1982,Bardeen1983}.

Now, a problem with $\Lambda$-CDM model is that the acceleration
of the Universe cannot be a permanent feature starting from the
Big-Bang. Because, an accelerating Universe is not favorable for
structure formation. This problem can be removed if one assumes
that the acceleration of the Universe is a recent phenomena. In
fact, some recent works\cite{Padmanabhan2002,Amendola2003} show
that the present accelerating phase was preceded by a decelerating
one and observational evidence\cite{Riess2001} also supports this
idea. The present work is done with this background in mind.

Phenomenological approach is one of the several ways of searching
such dark energy. In a recent work\cite{Ray2007a}, the equivalence
of three phenomenological variable $\Lambda$ models have been
shown. The behaviour of the same three forms of $\Lambda$ have
been studied when both $G$ and $\Lambda$ vary\cite{Ray2005}. But,
in both those works the equation of state parameter $\omega$ was
considered as a constant because, due to inability of current
observational data in separating a time-varying $\omega$ from a
constant one\cite{Kujat2002,Bartelmann2005}, in most of the cases
a constant value of $\omega$ is used. However, $\omega$, in
general, is a function of
time\cite{Chevron2000,Zhuravlev2001,Peebles2003}. It has already
been commented by Ray et al.\cite{Ray2007a} that for a more
accurate result, an investigation regarding time evolution of
$\omega$ may be taken up for searching better physical features.
In fact, the Statefinder diagnostic, used for distinguishing
various dark energy models, can be applied if the equation of
state of scalar potential has a direct relationship with the
Hubble parameter and its
derivative\cite{Starobinsky1998,Saini2000,Sahni2004}.

With those features of $q$ and $\omega$ in mind, an investigation
about the $\Lambda$-CDM Universe is done by selecting a specific
time-dependent form of $\Lambda$, viz., $\dot\Lambda\sim H^3$.
This particular time-varying $\Lambda$ model was studied by Reuter
and Wetterich\cite{Reuter1987} for finding a mechanism which would
explain the present small value of $\Lambda$ as a result of the
cosmic evolution. In the present work, the same $\Lambda$ model is
used for investigating a time evolving equation of state parameter
$\omega$ along with a possible signature flip of the deceleration
parameter $q$. This change of sign is very important for
$\Lambda$-CDM cosmology.

\section{Field Equations}
The Einstein field equations are given by
\begin{eqnarray}
R^{ij}-\frac{1}{2}Rg^{ij}= -8\pi G\left[T^{ij}-\frac{\Lambda}{8\pi
G}g^{ij}\right]
\end{eqnarray}
where the cosmological term $\Lambda$ is time-dependent, i.e.
$\Lambda = \Lambda(t)$ and $c$, the velocity of light in vacuum,
is assumed to be unity.

Let us consider the Robertson-Walker metric
\begin{eqnarray}
ds^2=
-dt^2+a(t)^2\left[\frac{dr^2}{1-kr^2}+r^2(d\theta^2+sin^2\theta
d\phi^2)\right]
\end{eqnarray}
where $k$, the curvature constant, assumes the values $-1$, $0$
and $+1$ for open, flat and closed models of the Universe
respectively and $a=a(t)$ is the scale factor. For the spherically
symmetric metric (2), field equations (1) yield Friedmann and
Raychaudhuri equations respectively given by
\begin{eqnarray}
3H^2+\frac{3k}{a^2}= 8\pi G\rho+\Lambda,
\end{eqnarray}
\begin{eqnarray}
3H^2+3\dot H= -4\pi G(\rho+3p)+\Lambda
\end{eqnarray}
where $G$, $\rho$ and $p$ are the gravitational constant, matter
energy density and pressure respectively and the Hubble parameter
$H$ is related to the scale factor by $H=\dot a/a$. In the present
work, $G$ is assumed to be constant. The generalized energy
conservation law for variable $G$ and $\Lambda$ is derived by
Shapiro et al.\cite{Shapiro2005} using Renormalization Group
Theory and also by Vereshchagin and
Yegorian\cite{Vereshchagin2006a} using a formula of Gurzadyan and
Xue\cite{Gurzadyan2003}. Vereshchagin and
Yegorian\cite{Vereshchagin2006b} have presented a phase portrait
analysis of the cosmological models relying on the Gurzadyan-Xue
type dark energy formula as mentioned above. A novel
interpretation of the physical nature of dark energy and
description of an internally consistent solution for the behavior
of dark energy as a function of redshift are provided by
Djorgovski and Gurzadyan\cite{Djorgovski2006} based on the vacuum
fluctuations model by Gurzadyan and Xue\cite{Gurzadyan2003}.

The conservation equation for variable $\Lambda$ and constant $G$
is a byproduct of the generalized conservation law and is given by
\begin{eqnarray}
\dot\rho+3(p+\rho)H= -\frac{\dot\Lambda}{8\pi G}.
\end{eqnarray}
Let us consider a relationship between the pressure and density of
the physical system in the form of the following barotropic
equation of state
\begin{eqnarray}
p= \omega\rho
\end{eqnarray}
where $\omega$ is the barotropic index which has been considered
here as time-dependent.

Using equation (6) we get from (5)
\begin{eqnarray}
8\pi G\dot\rho+\dot\Lambda= -24\pi G(1+\omega)\rho H.
\end{eqnarray}

Differentiating (3) with respect to $t$ we get for a flat Universe
($k=0$)
\begin{eqnarray}
-4\pi G\rho= \frac{\dot H}{1+\omega}.
\end{eqnarray}

As already mentioned in the introductory part, equivalence of
three phenomenological $\Lambda$-models (viz., $\Lambda \sim (\dot
a/a)^2$, $\Lambda \sim \ddot a/a$ and $\Lambda \sim \rho$) have
been studied in detail. So, similar type of variable-$\Lambda$
model may be investigated for a deeper understanding of both the
accelerating and decelerating phases of the Universe. Let us,
therefore, use the {\it ansatz}, $\dot\Lambda \propto H^3$, so
that
\begin{eqnarray}
\dot\Lambda= AH^3
\end{eqnarray}
where $A$ is a proportional constant.

Using equations (6), (8) and (9) we get from (4)
\begin{eqnarray}
\frac{2}{(1+\omega)H^3}\frac{d^2H}{dt^2}+\frac{6}{H^2}\frac{dH}{dt}=A.
\end{eqnarray}

If we put $dH/dt=P$, then equation (10) reduces to
\begin{eqnarray}
\frac{dP}{dH}+3(1+\omega)H= \frac{A(1+\omega)H^3}{2P}.
\end{eqnarray}

To arrive at fruitful conclusions, let us now solve equation (11)
under some specific assumptions.

\section{Solutions}
\subsection{$A=0$}
$A=0$ implies via equation (9), $\Lambda= constant$. In this case
equation (11) reduces to
\begin{eqnarray}
\frac{dP}{dH}+3(1+\omega)H= 0.
\end{eqnarray}
Solving equation (12) for $a(t)$, $\rho (t)$ and $H(t)$ we get
\begin{eqnarray}
a(t)= C_1t^{2/3(1+\omega)},
\end{eqnarray}
\begin{eqnarray}
H(t)= \frac{2}{3(1+\omega)}\frac{1}{t},
\end{eqnarray}
\begin{eqnarray}
\rho(t)= \frac{1}{6\pi G(1+\omega)^2}\frac{1}{t^2}
\end{eqnarray}
where $C_1$ is a constant.

It may be mentioned here that the above expressions for $a(t)$,
$H(t)$ and $\rho (t)$ can be recovered from the corresponding
expressions of Ray et al.\cite{Ray2007a} for $\alpha=0$, i.e.
$\Lambda=0$ where $\alpha$ is a parameter for the model $\Lambda
\sim H^2$ considered there. This means that the results (13), (14)
and (15) can be obtained either for constant $\Lambda$ (as in the
present case) or for null $\Lambda$ as in the case of Ray et
al.\cite{Ray2007a}. The essence of this is that, a null $\Lambda$
or constant $\Lambda$ will provide equivalent result. It may also
be mentioned here that by abandoning $\Lambda$, Einstein obtained
the expanding Universe while the same expanding Universe was
obtained by de Sitter for constant $\Lambda$.

Again, using equation (14) we can find the expression for the
deceleration parameter $q$ as
\begin{eqnarray}
q= -\left(1+\frac{\dot
H}{H^2}\right)=\left(\frac{1+3\omega}{2}\right).
\end{eqnarray}

From equation (16) we find that for an accelerating Universe,
$\omega<-1/3$. But, from equation (13)-(15) it is clear that
$\omega$ cannot be equal to $-1$. Moreover, the present value of
$q$ lies near $-0.5$\cite{Sahni1999} which can be obtained from
equation (16) by putting a value of $\omega$ which is equal to
$-2/3$. The sought for signature flipping of $q$ can be obtained
from equation (16) if one considers $\omega$ as time-dependent.

If $H_0$ and $t_0$ be the present values of $H$ and $t$, then from
equation (14) we can write,
\begin{eqnarray}
t_0= \frac{2}{3(1+\omega)H_0}.
\end{eqnarray}

Putting $\omega=-1/3$ in equation (17) and assuming $H_0=72$
kms$^{-1}$Mpc$^{-1}$ we find that the present age of the Universe
comes out as $13.58$ Gyr. which fits very well within the ranges
provided by various sources (for a list of data provided by
various sources one may consult Ray et al. \cite{Ray2007a}. In
this context it may be mentioned that for stiff-fluid ($\omega=1$)
Ray et al.\cite{Ray2007b} obtained the present age of the Universe
as $13.79$ Gyr. under the ansatz $\Lambda \sim H^2$.

\subsection{$1+\omega=-2P/H$}
By the use of the above substitution Eq. (11) becomes
\begin{eqnarray}
\frac{dP}{dH}-6P= -AH^2.
\end{eqnarray}

Solving equation (18) we get
\begin{eqnarray}
a(t)= C_2e^{-t/6}(sec Bt)^{1/6B},
\end{eqnarray}
\begin{eqnarray}
H(t)= \frac{1}{6} (tan Bt-1),
\end{eqnarray}
\begin{eqnarray}
\Lambda(t)= \frac{B}{6}\left[\frac{1}{2B} tan^2
Bt+\frac{2}{B}log(sec Bt)-\frac{3}{B}tan Bt+2t\right],
\end{eqnarray}
\begin{eqnarray}
\rho (t)= \frac{1}{48\pi G} (tan Bt -1),
\end{eqnarray}
\begin{eqnarray}
\omega (t)= -\left[1+\frac{2Bsec^2 Bt}{(tan Bt-1)}\right]
\end{eqnarray}
where $C_2$ is a constant and $B=A/36$.

For physically valid $H$, ~$tan Bt>1$. Then Eq. (23) implies that
for a positive $B$, $\omega$ must be less than $-1$ as in the case
of phantom energy. On the other hand if $B<0$, then $\omega$ can
be greater than $-1$ as well.

Similar type of trigonometric solutions have been obtained by
Mukhopadhyay and Ray\cite{Mukhopadhyay2005} for polytropic
equation of state with constant $\omega$ in non-dust case under
the {\it ansatz} $\Lambda \sim \ddot a/a$. Simple trigonometric
solution for the scale factor was also obtained by Banerjee and
Das\cite{Banerjee2005} in scalar field model. But, they obtained
their solution by making a special assumption on the deceleration
parameter while the present solution is a result of a supposition
on the equation of state parameter $\omega$.

Again, using equation (20) we get
\begin{eqnarray}
q= -\left[1+\frac{6B sec^2 Bt}{(tan Bt-1)^2}\right].
\end{eqnarray}

From equation (24) we find that, a signature flipping of $q$ is
possible if $B<0$. So, the merit of this case lies the fact that
the same change of sign of $q$ can be obtained here by using a
time-dependent form of $\omega$ and not making any special
assumption on $q$ directly as was done by Banerjee and
Das\cite{Banerjee2005}. This once again shows that the equation of
state parameter is a key ingredient of cosmic evolution.

\subsection{$1+\omega= -P/3H^2$}
With the above assumption, Eq. (11) becomes
\begin{eqnarray}
\frac{dP}{dH}-\frac{P}{H}= -\frac{A}{6}H.
\end{eqnarray}
Solving equation (25) we get our solution set as
\begin{eqnarray}
a(t)= C_4t^{6/A},
\end{eqnarray}
\begin{eqnarray}
H(t)= \frac{6}{At},
\end{eqnarray}
\begin{eqnarray}
\rho(t)= \frac{27}{\pi G A^2}\frac{1}{t^2},
\end{eqnarray}
\begin{eqnarray}
\Lambda(t)= -\frac{108}{A^2}\frac{1}{t^2},
\end{eqnarray}
\begin{eqnarray}
\omega(t)= \frac{A}{18} -1
\end{eqnarray}
where $C_4$ is an integration constant.

Thus, we find that in this case the scale factor admits a power
law solution, $H$ varies inversely as $t$ and $\rho$ as well as
$\Lambda$ follow the well known inverse square law with $t$. This
type of solution was obtained by Ray et al.\cite{Ray2007a} for
$\Lambda \sim (\dot a/a)^2$, $\Lambda \sim \ddot a/a$ and $\Lambda
\sim \rho$ models. For physical validity $A>0$. But, in this case
$\Lambda <0$ for real $A$ and hence represents an attractive
force. However, $\Lambda$ can be a repulsive force as well if $A$
is a complex number. Now, a complex $A$ means a complex scale
factor. So, this particular case can be thought of as a
phenomenological version of spintessence model of Banerjee and
Das\cite{Banerjee2006} where a complex scalar field of the form
$\phi=e^{i\omega t}$ is used to search for the cosmic
acceleration. But, in that case $\omega$ is a constant. Also, for
$A=6$ if we take the present value of the Hubble parameter as
$H_0=72$ kms$^{-1}$Mpc$^{-1}$ then, from equation (27) the present
age of the Universe comes out as $13.58$ Gyr. which agrees very
well with the estimated value\cite{Ray2007a}. Now, for $A=6$ we
have $\omega=-2/3$ and the scale factor grows linearly with time.
Again, in this case
\begin{eqnarray}
q= -\left[1+\frac{A}{6}\right].
\end{eqnarray}
Equation (31) shows that for a positive $A$, the Universe expands
with a constant acceleration. For $A=6$ the amount of acceleration
is $-2$. So, in this case of the phenomenological model
$\dot\Lambda\sim H^3$, the deceleration parameter $q$ does not
show any change in sign during cosmic evolution.

\section{Discussions}
The main objectives of the present work were to search for a
signature flip of $q$ and to find time-dependent expression for
the equation of state parameter. In that respect, this work in
general has fulfilled its goal. By selecting a time dependent form
of the cosmological parameter $\Lambda$, through some analytical
study, it has been possible to show that a change in sign of the
deceleration parameter can be achieved under some special
assumptions (Sec. 3.2). It has also been possible to derive time
dependent expressions for the equation of state parameter
$\omega$. It is found that $\omega$ can be less than $-1$ as well
which is compatible with SN Ia data\cite{Knop2003} and SN Ia data
with CMB anisotropy and galaxy-cluster
statistics\cite{Tegmark2004b}.

It would be worthwhile to mention here that in the present work
the parameter $A$ plays a vital role for studying the cosmic
evolution of various phases of the Universe. For instance, a null
$A$ presents us a case of constant $\Lambda$ (Sec. 3.1) whereas
positive and negative $A$ show the possibility of $\omega<-1$
(phantom energy) or $\omega>-1$ respectively (Sec. 3.2). The well
deserved signature flipping of $q$ is also possible for $A<0$
(Sec. 3.2). A possibility of a complex $A$ is provided in Sec. 3.3
which has similarity with the work of Banerjee and
Das\cite{Banerjee2006}. For a specific value of $A$ the age of the
Universe and the value of $q$ are calculated also (Sec. 3.3). It
is interesting to note that similar type of case study for the
cosmic evolution has been done by
Khachatryan\cite{Khachatryan2007} using a parameter $b$ for null,
positive and negative values of it. So, whether there exists any
internal physical relationship between the present work and that
of Khachatryan\cite{Khachatryan2007} may be a subject matter of
future investigation.

Determination of the present value of the Hubble parameter through
analysis of CMB data from WMAP and HST Key Project suggests that
value of the equation of state parameter for dark energy models
should be less than $-0.5$ at the $95 \%$ confidence
level\cite{Spergel2003}. So, in some cases a Big Rip may not be
impossible. For $\Lambda$-CDM models, the Statefinder diagnostic
satisfies the condition ${\frac{d\ddot a}{dt}}/aH^3=1$. Since
${\frac{d\ddot a}{dt}}/a$ can be expressed in terms of $H$, $\dot
H$ and $\ddot H$, so it is easy to verify that first (Sec. 3.1)
and third (Sec. 3.3) cases satisfy the above condition prescribed
by the Statefinder diagnostic for $\omega=0$ and $A=9$
respectively.

However, the $\Lambda$-CDM Universe with $\omega=-1$, where the
sine hyperbolic form of the scale factor can reflect both matter
dominated past and accelerated expansion in
future\cite{Sahni2000}, can not be achieved through this model.
Equation (11) shows that for $\omega=-1$, $H$ grows linearly with
time which does not fit with the present cosmological scenario.
However, through the present model, it has been possible to
provide some interesting situations which were obtained earlier by
different researchers and are already discussed in respective
Sections. Some awkward cases, such as constant energy density
(which can be obtained by putting $1+\omega=2P$ in equation (11))
can be found in the work of Ray\cite{Ray2006} in relation to
electromagnetic mass in $n+2$ dimensional space-time. Some other
works\cite{Bonnor1960,Wilson1968,Cohen1969,Florides1983,Gron1985,Ivanov2002}
also admit constant matter distribution in their solutions.
Finally, it should be mentioned that the present work is done
making $\Lambda$ variable and keeping $G$ constant. So, it may be
an interesting study when the present model is combined with a
variable $G$ and obey the generalized energy conservation law
derived by Shapiro et al.\cite{Shapiro2005} and Vereshchagin and
Yegorian\cite{Vereshchagin2006a}. That can be a subject matter of
our future investigation.

\section*{Acknowledgments}
One of the authors (SR) is thankful to the authority of
Inter-University Centre for Astronomy and Astrophysics, Pune,
India, for providing Associateship programme under which a part of
this work was carried out.




\end{document}